\documentclass[pra, aps, superscriptaddress, twocolumn, showpacs, 10pt, floatfix]{revtex4-1}
\usepackage{array}
\usepackage{graphicx}
\usepackage{amsmath}
\usepackage{amssymb}
\newcommand{\opt}{\mathrm{opt}}
\newcommand{\SPD}{\text{SPD}}
\newcommand{\ThD}{\text{ThD}}
\begin{document}

\title{Single-photon sources based on asymmetric spatial multiplexing with optimized inputs}

\author{Peter Adam}
\email{adam.peter@wigner.hu}
\affiliation{Institute of Physics, University of P\'ecs, Ifj\'us\'ag \'utja 6, H-7624 P\'ecs, Hungary}
\affiliation{Institute for Solid State Physics and Optics, Wigner Research Centre for Physics,\\ P.O. Box 49, H-1525 Budapest, Hungary}
\author{Ferenc Bodog}
\affiliation{MTA-PTE High-Field Terahertz Research Group H-7624 P\'ecs, Hungary}
\author{Matyas Koniorczyk}
\affiliation{Institute for Solid State Physics and Optics, Wigner Research Centre for Physics,\\ P.O. Box 49, H-1525 Budapest, Hungary}
\author{Matyas Mechler}
\affiliation{Institute of Physics, University of P\'ecs, Ifj\'us\'ag \'utja 6, H-7624 P\'ecs, Hungary}

\begin{abstract}
We develop a statistical theory describing the operation of multiplexed single-photon sources equipped with photon-number-resolving detectors that includes the potential use of different input mean photon numbers in each of the multiplexed units.
This theory accounts for all relevant loss mechanisms and allows for the maximization of the single-photon probabilities under realistic conditions by optimizing the different input mean photon numbers unit-wise and the detection strategy that can be defined in terms of actual detected photon numbers.
We apply this novel description to analyze periodic single-photon sources based on asymmetric spatial multiplexing realized with general asymmetric routers.
We show that optimizing the different input mean photon numbers results in maximal single-photon probabilities higher than those achieved by using optimal identical input mean photon numbers in this setup.
We identify the parameter ranges of the system for which the enhancement in the single-photon probability for the various detection strategies is relevant.
An additional advantage of the unit-wise optimization of the input mean photon numbers is that it can result in the decrease of the optimal system size needed to maximize the single-photon probability.
We find that the highest single-photon probability that our scheme can achieve in principle when realized with state-of-the-art bulk optical elements is 0.935. This is the highest one to our knowledge that has been reported thus far in the literature for experimentally realizable single-photon sources.
\end{abstract}

\maketitle

\section{Introduction}
Single-photon sources (SPS) are the key elements in numerous applications in the field of quantum information processing~\cite{Meyer2020,Knill2001, Kok2007, Gisin2002, Scarani2009, Duan2001, Sangouard2011, Bennett1993, Bouwmeester1997, Merali2011, Koniorczyk2011, Spring2013, Broome2013, Tillmann2013} and photonic quantum technology~\cite{Gerry1999, Lund2004, He2009, Adam2010, Lee2012}.
Heralded single-photon sources (HSPS) based on spontaneous parametric down-conversion (SPDC) \cite{Fiorentino2007, Mosley2008, Zhong2009, Evans2010, Eckstein2011, Brida2011, Broome2011, Horn2013} and spontaneous four-wave mixing (SFWM) \cite{Smith2009, Cohen2009, Soller2011, MScott2015} are promising candidates for realizing periodic SPS.
In such sources the detection of one member (called the idler) of a correlated photon pair generated in these nonlinear processes heralds the presence of its twin photon (termed as the signal).

Though HSPS can yield highly indistinguishable single photons \cite{Mosley2008,Evans2010, Eckstein2011, Fortsch2013} that are required for applications, the generation of photon pairs in such nonlinear sources is probabilistic by nature posing a limit on the achievable single-photon probability.
To overcome the problem of multiphoton events that occasionally occur in the pair generation, various techniques of multiplexing, namely spatial multiplexing \cite{Migdall2002, ShapiroWong2007, Ma2010, Collins2013, Meany2014, Francis2016, KiyoharaOE2016} and time multiplexing \cite{Pittman2002, Jeffrey2004, Mower2011, Schmiegelow2014, Francis2015, Kaneda2015, Rohde2015, XiongNC2016, Hoggarth2017, HeuckNJP2018, Kaneda2019, Lee2019, MagnoniQIP2019} were proposed in the literature.
In multiplexed SPS, heralded photons generated in a set of multiplexed units realized in time or in space are rerouted to a single output mode by a switching network.
In order to suppress the multi-photon noise, the mean photon number of the generated photon pairs should be kept low in a multiplexed unit, while the use of several multiplexed units guarantees the high probability of successful heralding in all the multiplexed units so that high single-photon probabilities can be achieved.

In the case of time multiplexing, idler photons originating from a nonlinear photon pair source are detected in well-defined time slots within the planned time period.
After a successful detection event the heralded signal photons are delayed to leave the system at the end of this period. The necessary delay can be realized by using a binary division strategy \cite{Mower2011,Schmiegelow2014,MagnoniQIP2019, Lee2019}, or by a switchable optical storage cavity or loop \cite{Pittman2002, Jeffrey2004, Rohde2015, HeuckNJP2018}.

The idea of spatial multiplexing is based on using several individual pulsed HSPS in parallel.
After a detection event in the idler arm of a photon pair source, a set of binary photon routers (2-to-1) is used to direct the corresponding heralded signal photons to a single output.
The routers can be arranged into a symmetric (binary tree) or asymmetric (chain) structure \cite{MazzarellaPRA2013,BonneauNJP2015}.

Experimental realizations of spatial multiplexing have been reported up to four multiplexed units by using SPDC in bulk crystals \cite{Ma2010,KiyoharaOE2016} and waveguides \cite{Meany2014}, and by using SFWM up to two multiplexed units in photonic crystal fibers \cite{Collins2013, Francis2016}.
As of time multiplexing, delay-loop-based arrangements have been realized up to four time slots in fiber-based systems with SFWM \cite{XiongNC2016,Hoggarth2017}.
SPS were realized in experiments using optical storage cavities and SPDC sources via large-scale time multiplexing up to 40 time slots \cite{Kaneda2015, Kaneda2019}.

In principle, the increase of the number of multiplexed units and the simultaneous decrease of the mean photon number of the photon pairs in an ideal lossless multiplexed system lead the single-photon probability to tend to one asymptotically. This does not hold in real multiplexed systems as the losses of the optical elements in both the heralding stage and the multiplexing system limit the performance of multiplexed SPS \cite{MazzarellaPRA2013, BonneauNJP2015}. In order to analyze such systems, a full statistical theory has been developed that takes all relevant loss mechanisms into account \cite{Adam2014,Bodog2016,Bodog2020}.
Such a theory was developed in Ref.~\cite{Adam2014} for the description and optimization of both spatially and time-multiplexed systems using threshold detectors.
This theoretical framework was extended in Ref.~\cite{Bodog2016} to describe SPS based on combined spatial and time multiplexing in a single setup \cite{GlebovAPL2013, LatypovJPCS2015, MendozaOptica2016}.
The theory in Ref.~\cite{Adam2014} was further generalized in Ref.~\cite{Bodog2020} to cover the case of multiplexed SPS operated with photon-number-resolving detectors (PNRDs).

Using PNRDs in multiplexed SPS enables the application of a broad range of detection strategies: detecting suitably chosen sets of predefined number of photons in the idler arm for which the generated signal photons are allowed to enter the multiplexer.
Allowing more than one photons to enter into the multiplexing system can be advantageous for multiplexers having higher losses, as it has been shown in Ref.~\cite{Bodog2016}.
Single-photon detectors with photon-number-resolving capabilities were already used in recent multiplexed periodic SPS experiments \cite{Collins2013, Kaneda2015, Francis2016, KiyoharaOE2016, XiongNC2016, Hoggarth2017, Kaneda2019} in order to avoid the occurrence of multi-photon events at the heralding stage of these systems.
Meanwhile, various realizations of high-efficiency inherent PNRDs have been developed including  transition edge sensors \cite{Cabrera1998, Miller2003, Rosenberg2005, Lita2008, Lita2009, Lita2010, Fukuda2011, Schmidt2018, Fukuda2019}, quantum dot optically gated field-effect transistors \cite{Gansen2007, Kardynal2007}, superconducting nanowire detectors~\cite{Divochiy2008, Jahanmirinejad2012}, and fast-gated avalanche photodiodes~\cite{Kardynal2008, Thomas2010}. This progress motivates the development of theoretical frameworks for multiplexed SPS equipped with PNRDs.

The significance of  full statistical treatments is that they make it possible to optimize SPS: to determine, for a given set of loss parameters, the optimal system size and a mean photon number of the photon pairs in the multiplexed units for which the output single-photon probability is maximal.
In Ref.~\cite{Bodog2020} the optimization results for SPS based on symmetric spatial multiplexing realized with symmetric routers,  binary bulk time multiplexing, and storage loop time multiplexing were presented; in all of these cases the results for either single-photon detectors or threshold detectors were discussed.

In Ref.~\cite{MazzarellaPRA2013} an asymmetric architecture was proposed for the spatially multiplexed SPS instead of the symmetric structure discussed previously in the literature.
In this novel architecture the constituent routers are arranged into a chain-like structure instead of the binary-tree-like structure of the symmetric arrangement.
A special property of these asymmetric architectures is that the single-photon probability increases with the increase of the number of routers, provided that on multiple detection events the arm with the smallest total transmission coefficient is chosen.
In this respect this novel structure is similar to the time multiplexing schemes based on a switchable optical storage cavity or loop operated with a similar logic.
The idea of using increasing mean photon numbers of photon pairs in the subsequent multiplexing units instead of identical ones was also proposed in the cited paper in order to compensate for the increasing losses, that is, for the decreasing total transmission coefficients characterizing the subsequent arms in the chain structure.
However, no detailed analysis of the question was performed. Instead of optimizing the mean photon numbers unit-wise, only a simple scaling function was used to address this problem. Besides, the analysis was restricted to SPS based on asymmetric spatial multiplexing realized with symmetric routers and threshold detectors.

In this paper we develop the statistical theory for multiplexed SPS equipped with PNRDs that incorporates the use of different mean photon numbers of the photon pairs at the different multiplexed units.
This theory, which is a generalization of the previous statistical model introduced in \cite{Bodog2020}, includes all relevant loss mechanisms and allows for the maximization of the single-photon probabilities under realistic conditions by optimizing the different mean photon numbers unit-wise.
We apply our novel description to analyze SPS based on general asymmetric spatial multiplexing realized with asymmetric routers and PNRDs.
As an appropriate tool to determine the optimal different mean photon numbers at the different multiplexed units, we apply a genetic algorithm.
We show that, in principle, this scheme can produce the highest single-photon probability when realized with state-of-the-art bulk optical elements.

This paper is organized as follows: in Sec.~\ref{sec:ASM} we describe the system realizing asymmetric spatial multiplexing with asymmetric routers. In Sec.~\ref{Sec:math} we introduce the theoretical framework for multiplexed SPS equipped with PNRDs incorporating the use of different mean photon numbers at the different multiplexed units that can be applied to perform the optimization and analysis of such systems. The results of the optimization for asymmetric spatial multiplexing are presented in Sec.~\ref{Sec:result}. Finally, conclusions are drawn in Sec.~\ref{Sec:conc}.

\section{Asymmetric spatial multiplexing}\label{sec:ASM}

\begin{figure*}[bt]
    \centering
    \includegraphics[width=0.45\textwidth]{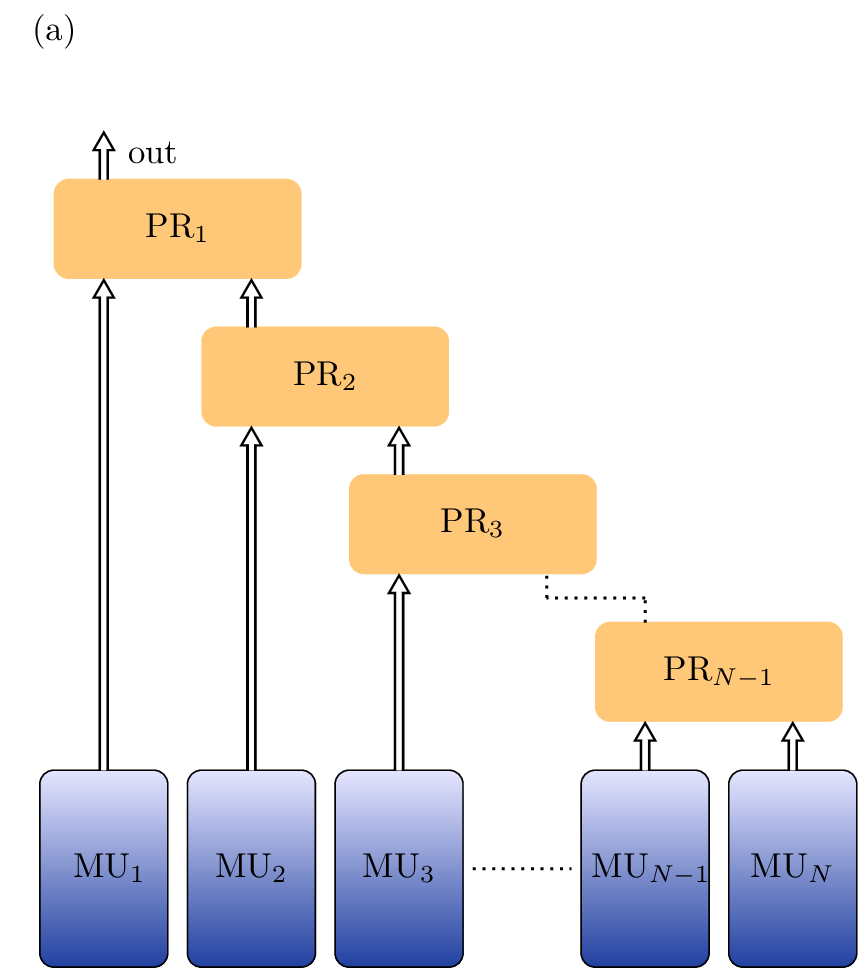}\hspace{1cm}
    \includegraphics[width=0.3\textwidth]{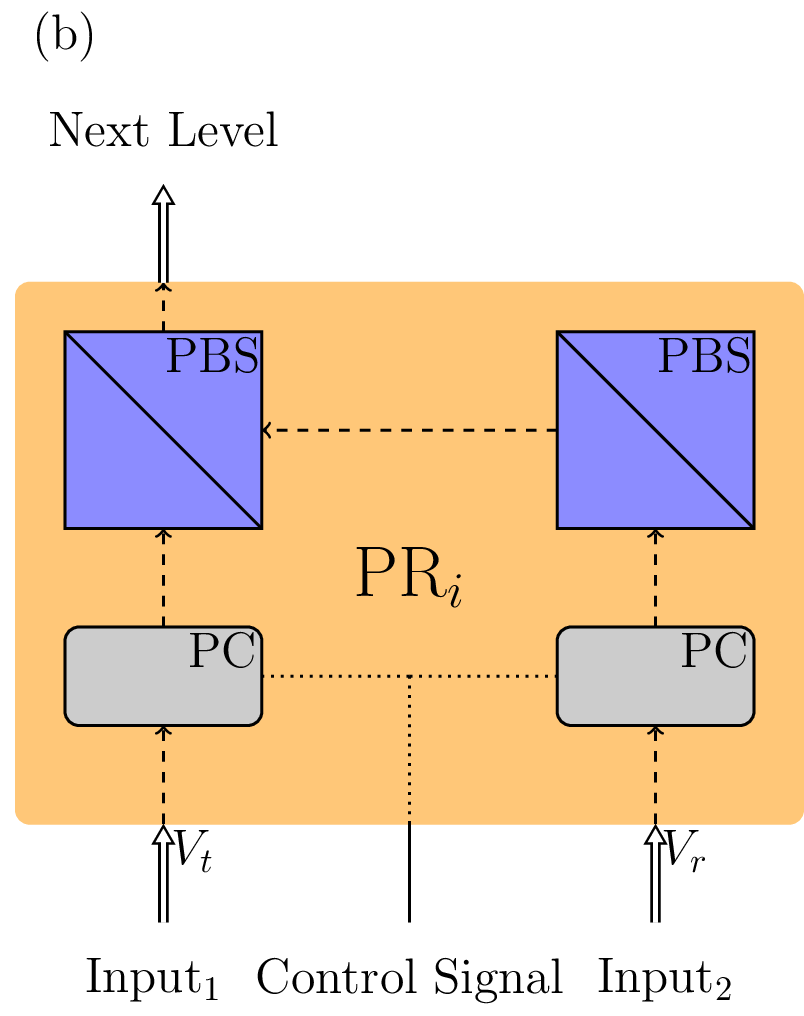}
\caption{(a) Schematic figure of a periodic single-photon source based on asymmetric spatial multiplexing. MU$_i$s denote multiplexed units and PR$_i$s denote 2-to-1 photon routers. (b) Scheme of the bulk optical photon routers PR$_i$. PCs denote Pockels cells, PBSs denote polarizing beam splitters. $V_t$ and $V_r$ denote transmission coefficients that characterize the losses for Input$_1$ and Input$_2$, respectively.
\label{fig:1}}
\end{figure*}

The idea of asymmetric spatial multiplexing is to reroute heralded photons generated in a set of multiplexed units by a chained architecture of photon routers \cite{MazzarellaPRA2013}.
The schematics of this multiplexing is presented in Fig.~\ref{fig:1}(a).
In the figure the MU$_i$-s ($i=1,\dots,N$) denote multiplexed units.
Each of them contains a nonlinear photon pair source, a detector placed in the path of the idler mode and a delay line optionally placed in the path of the signal mode of the source.
The delay line can be needed to ensure enough time for the operation of the necessary control logic governing the routers. A signal photon of a pair generated by a nonlinear source in a multiplexed unit is allowed to enter the multiplexer provided that its twin, the idler photon is detected.
The idler photon can be detected either with a threshold detector, or a single-photon detector or, generally, a photon-number-resolving detector (PNRD) with arbitrary detection strategy.
The periodicity of this single-photon source is ensured by pulsed pumping of the nonlinear photon pair sources.

In the studied setup the $N-1$ photon routers (PRs) forming the asymmetric spatial multiplexer are assumed to be identical.
The binary routers used in spatial multiplexers are usually considered to be symmetric, but it is not necessarily the case.
Hence we consider general, asymmetric routers in our analysis. One possible bulk optical realization of an asymmetric router is presented in Fig.~\ref{fig:1}(b).
This router has two input ports, a single output, and it contains two Pockels-cells (PC) and two polarizing beam splitters (PBS).
As the signal photons are generated with a known polarization, a priority logic can control the PCs so that a single input mode reaches the output of the router.
The chosen mode is selected by the PBSs according to the polarization set by the control logic via the PCs.
If signal photons are heralded in more than one multiplexed units, a reasonable choice for the control logic is to direct photons to the output from the multiplexed unit having the smallest loss, that is, from the one closest to the output of the multiplexer.
In the arrangement presented in Fig.~\ref{fig:1}(a) this is the unit with the smallest index.

The operation of an asymmetric router can be characterized with two transmission coefficients $V_t$ and $V_r$ corresponding to the transmissions of the photons entering the router at Input$_1$ and Input$_2$, respectively. In the case of the bulk optical asymmetric router presented in Fig.~\ref{fig:1}(b) the transmission $V_t$ quantifies the losses due to the transmission through a PC and a PBS and the possible propagation loss in the router. The transmission $V_r$ describes the losses introduced by the transmission through a PC and the two reflections in the PBSs, and the possible propagation loss in the router. We refer to the coefficients $V_t$ and $V_r$ as the transmission and reflection efficiencies in what follows.

The total transmission probability $V_n$ of the $n$-th arm of the asymmetric spatial multiplexer built from the proposed bulk optical asymmetric routers reads
\begin{equation}
\begin{alignedat}{3}
& V_n = V_bV_tV_r^{n-1} \quad && \text{if} \quad && n<N, \\ 
& V_n = V_bV_r^{n-1} \quad &&  \text{if} \quad &&  n=N,\label{eq:ASM}
\end{alignedat}
\end{equation}
where $N$ is the number of multiplexed units.
The parameter $V_b$ is a general transmission coefficient characterizing all the losses experienced by the heralded photons while they propagate to the input of the multiplexer, such as the losses of the involved passive optical elements.  
The presented spatially multiplexed SPS can be operated assuming identical values of the mean number of the generated photon pairs $\lambda$ in each of the multiplexed units.
Choosing different mean photon numbers $\lambda_n$ in each of the multiplexed units as it has been suggested in Ref.~\cite{MazzarellaPRA2013} is, however, a reasonable way to compensate for the effect of the different total transmission efficiencies $V_n$ of the subsequent arms of the multiplexer.
The use of different mean photon numbers can be implemented in experiments by the proper adjustment of the pumps of the nonlinear photon pair sources in each of the multiplexed units.

\section{Statistical framework}\label{Sec:math}

In Ref.~\cite{Bodog2020} we have introduced a full statistical framework capable of describing spatially or time multiplexed periodic SPS equipped with PNRDs. In the following we generalize this framework to incorporate the possibility of using different mean photon numbers $\lambda_n$ in each of the multiplexed units.

Let us consider a single-photon source containing $N$ multiplexed units.
Assume that the nonlinear source in the $n$-th multiplexed unit generates $l$ photon pairs and the detection of a predefined number of photons $j$ ($j \leq l$) by a PNRD during a heralding event causes the opening of the corresponding input port of the multiplexer. Under these assumptions one can express the probability that the output of the multiplexer is reached by $i$ signal photons as
\begin{widetext}
\begin{align}
\begin{split}
P_i^{(S)}=
\prod_{k=1}^{N}\big(
1-\sum_{j\in S}P_k^{(D)}(j)
\big)\delta_{i,0}+\sum_{n=1}^N
\left[
\prod_{k=1}^{n-1}
\big(1-\sum_{j\in S}P_k^{(D)}(j)\big)^{(1-\delta_{1,n})}\sum_{l=i}^\infty\sum_{j\in S} P^{(D)}(j|l)P_n^{(\lambda_n)}(l)V_n(i|l)
\right].
\label{general_formula}  
\end{split}
\end{align}
\end{widetext}
In this formula the 
probability of detecting exactly $j$ photons in the $n$th multiplexed unit is denoted by $P_n^{(D)}(j)$, while $P^{(D)}(j|l)$ stands for the conditional probability of registering $j$ photons provided that $l$ photons arrive at the detector, and $P_n^{(\lambda_n)}(l)$ denotes the probability of generating $l$ photon pairs in the $n$th multiplexed unit when the mean photon number of the generated photon pairs is $\lambda_n$ in that unit.
$V_n(i|l)$ is the conditional probability of the event that the output of the multiplexer is reached by $i$ photons provided that the number of signal photons arriving from the $n$th multiplexed unit into the system is $l$.

Equation~\eqref{general_formula} contains a summation over the elements of the set $S$: this set comprises the predetermined number of detected photons in a single multiplexed unit for which the generated signal photons are allowed to enter the multiplexer.
Therefore this set describes the application of an optional detection strategy that can be realized only by PNRDs.
The set $S$ can contain any combination of the elements of the set of positive integers $\mathbb{Z}^+$ up to a predefined value $J_b$ determined by the applied PNRD. This number corresponds to the maximum number of detected photons that can be distinguished by the detector.
The case $S=\mathbb{Z}^+$ corresponds to ignoring the number of detected photons when the PNRD actually acts as a threshold detector.
When the mean photon numbers $\lambda_n$ are chosen to be the same for all the multiplexed units (that is, $\lambda_n=\lambda$) Eq.~\eqref{general_formula} reduces to the respective formula of Ref.~\cite{Bodog2020}.

Equation \eqref{general_formula} comprises two terms. The first term corresponds to the case when the photon number registered by the detectors in the multiplexed system is not in the set $S$.
As in this case no photon enters the multiplexed system, this term contributes to the probability $P_0^{(S)}$ only, describing the case in which no photon reaches the output. The second term describes the case when, even though there are $l$ photons entering the multiplexer from the $n$-th multiplexed unit after heralding, only $i$ of these reach the output due to the losses of the multiplexer.
In this second term the product expresses the fact that no photon arrives in the first $n-1$ multiplexed units.

Assuming that the PNRD has a detector efficiency $V_D$, the conditional probability $P^{(D)}(j|l)$ describing the case when the detector detects $j$ out of $l$ photons ($j \leq l$) in a multiplexed unit in the second term of Eq.~\eqref{general_formula} reads
\begin{eqnarray}
P^{(D)}(j|l)=\binom{l}{j}V_D^j(1-V_D)^{l-j}.
\end{eqnarray}
Then the total probability $P_n^{(D)}(j)$ of detecting $j$ photons in the $n$-th multiplexed unit can be written as
\begin{eqnarray}
P_n^{(D)}(j)=\sum_{l=j}^\infty P^{(D)}(j|l)P_n^{(\lambda_n)}(l).
\end{eqnarray}
We note that beside the finite detector efficiency $V_D$ our analysis does not take into account other possible detector imperfections such as dark counts and the miscategorization of the actual photon count values of PNRDs.
This does not pose any significant limitation against the realistic nature of our model, consult Ref.~\cite{Bodog2020} for a detailed justification.

We assume that the probability distribution of pair generation $P_n^{(\lambda_n)}(l)$ is Poissonian, that is,
\begin{equation}
P_n^{(\lambda_n)}(l)=\frac{\lambda_n^l e^{-\lambda_n}}{l!},
\end{equation}
where $\lambda_n$ is the mean photon number of the photon pairs generated in the $n$th multiplexed unit. This quantity is the input of the heralding process, therefore we use the term \emph{input mean photon number} to refer to it in the following.
The number of the generated photon pairs follows this statistics in the case of multimode SPDC or SFWM processes, that is, when weaker spectral filtering is applied in the system~\cite{Avenhaus2008, Mauerer2009, Takesue2010, Almeida2012, Collins2013, Harder2016, KiyoharaOE2016}.
This assumption on the distribution enables us to compare the results with those presented in a significant part of the experimental and theoretical literature related to SPS, which were also achieved by considering the Poissonian distribution~\cite{Jeffrey2004, ShapiroWong2007, Ma2010, Mower2011, MazzarellaPRA2013, Adam2014, Bodog2016, MagnoniQIP2019}.
We note that Eq.~\eqref{general_formula} remains valid for any input distributions such as thermal distribution that occurs for photon pairs produced in single-mode, that is, spectrally narrow-filtered SPDC or SFWM processes \cite{Avenhaus2008, Mauerer2009, Takesue2010, Almeida2012, Collins2013, Harder2016, KiyoharaOE2016}.

The conditional probability $V_n(i|l)$ of the event that $i$ signal photons reach the output of the multiplexer given that $l$ signal photons enter the multiplexer at the $n$-th multiplexed unit can be calculated as
\begin{equation}
V_n(i|l)=\binom{l}{i}V_n^i(1-V_n)^{l-i},
\end{equation}
where $V_n$ is the total probability of transmission of the $n$-th arm of the multiplexer.
For the considered setup of Fig.~\ref{fig:1} $V_n$ is defined in Eq.~\eqref{eq:ASM}.
Finally, we remark that the statistical model in Eq.~\eqref{general_formula} can describe both spatial and time multiplexing with different input mean photon numbers of the photon pairs in the multiplexed units, though in the case of time multiplexers the realization of such an input is not trivial.

\section{Optimized SPS based on asymmetric spatial multiplexing} \label{Sec:result}
In this Section we present our results regarding the optimization of the SPS based on asymmetric spatial multiplexing described in Sec.~\ref{sec:ASM}.
Our aim is to analyze the system for experimentally feasible loss parameters.
Accordingly, the upper boundaries of the considered ranges of the loss parameters in our model are determined by assuming optical elements in the system with the best parameters available with state-of-the-art technology.
Polarizing beam splitters with 99.9\% reflectivity for S-polarized light and 99.3\% transmittivity for P-polarized light were already used in Ref.~\cite{Peters2006}.
Pockels cells with 99.2\% transmission were reported in Ref.~\cite{Kaneda2019}.
Taking into account these parameters and that the propagation losses in the bulk optical router in Fig.~\ref{fig:1}(b) can be very low, the highest values of the reflection and transmission efficiencies of the router are chosen to be $V_r=0.99$ and $V_t=0.985$, respectively.
Detector efficiencies as high as $V_D=0.98$ have already been reported with almost ideal photon number discrimination at low photon numbers using transition edge sensors in the near-infrared regime~\cite{Fukuda2011}.
The general transmission coefficient $V_b$ strongly depends on actual experimental realization of the system; we assume the value of $V_b=0.98$ for its highest feasible value.
The lower boundaries of the ranges of the loss parameters are chosen so that the characteristic behavior of the system can be revealed, hence we set the lower boundaries for the coefficients $V_r$, $V_D$, and $V_b$ to $0.8$.
Let us note that in the proposed setup the total transmission coefficient $V_n$ does not scale with the transmission efficiency $V_t$, its role in the first $n-1$ terms is similar to the role of the general transmission coefficient $V_b$. For this reason, we fix its value to be $V_t=0.985$ in all calculations.

The optimization procedure within the statistical framework presented in Sec.~\ref{Sec:math} can be accomplished in the following way. We fix the reflection and transmission efficiencies $V_r$ and $V_t$, the general transmission coefficient $V_b$, and the detector efficiency $V_D$. We fix the detection strategy by defining the set $S$ as well. We consider sets which contain numbers from 1 up to $J\leq J_b$. Thus the maximum accepted photon number $J$ exactly defines the set $S$.

The next step is to determine the different optimal input mean photon numbers $\lambda_{n,\opt}$ for each of the sequential numbers $n$ identifying the multiplexed units up to an overall number of multiplexed units $N$ that yields the highest single-photon probability $P_{1,N}^{S,\lambda_{n}}$ achievable for the given $N$ adopting the detection strategy defined by $S$. This procedure must be performed from $N=1$ up to a high enough value whose choice will be clarified in detail later. This task can be accomplished by using a genetic algorithm \cite{Goldberg1989, MATLAB}.

A genetic algorithm is a randomized search algorithm that can be used for solving both constrained and unconstrained optimization problems \cite{Goldberg1989}. In this multistep method to every step (or generation) a number of random points (called population) are assigned in the parameter space and a set of the best values of the objective function (also termed as the fitness function) are selected around which a new population is chosen. This procedure is repeated until a certain constraint is fulfilled.

\begin{figure*}[tb]
    \centering
    \includegraphics[width=0.99\columnwidth]{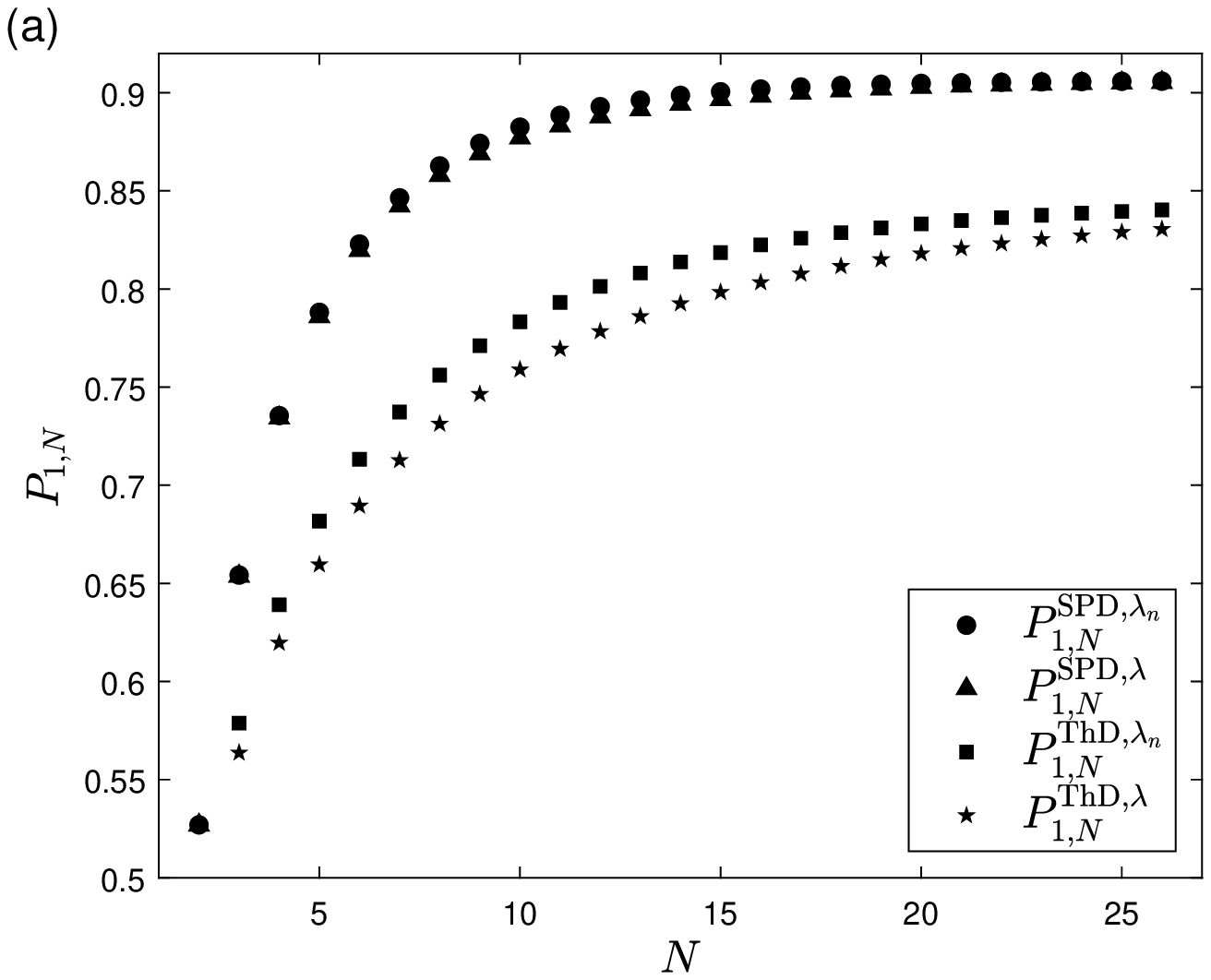}\hfill
    \includegraphics[width=0.99\columnwidth]{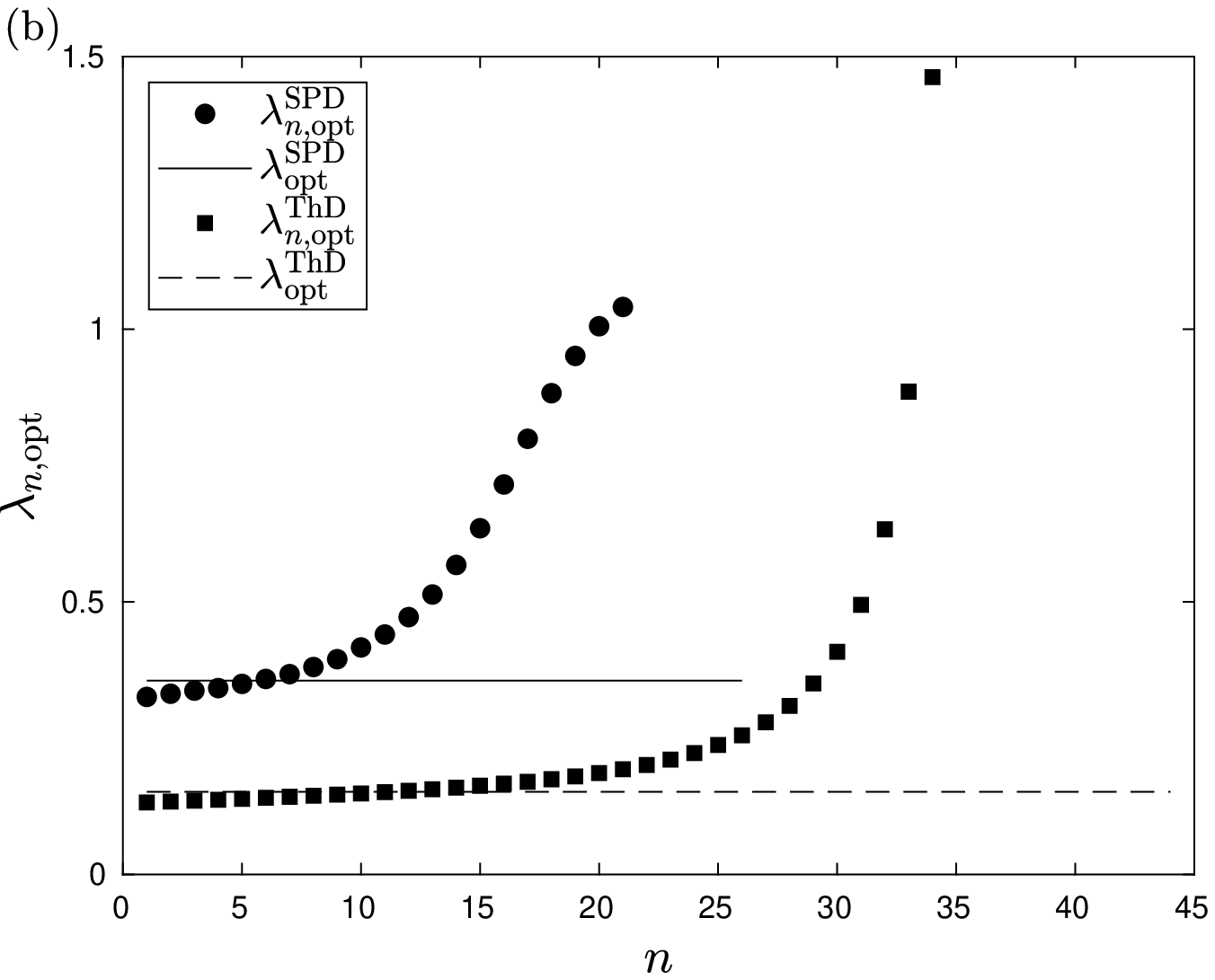}
\caption{(a) Achievable single-photon probabilities $P_{1,N}$ as a function of the number of multiplexed units $N$ for the SPD and ThD strategies assuming different and identical optimal input mean photon numbers $\lambda_{n,\opt}$ and $\lambda_{\opt}$, respectively, for each of the multiplexed units for the reflection efficiency $V_r=0.99$, the general transmission coefficient $V_b=0.98$, and the detector efficiency $V_D=0.9$. (b) Optimal input mean photon numbers $\lambda_{n,\opt}$ as a function of the sequential number $n$ of the multiplexed units for the considered detection strategies up to the corresponding optimal numbers of multiplexed units $N_\opt$ and for the parameters of Fig.~\ref{fig:2}(a). The values of the identical optimal input mean photon numbers $\lambda_{\opt}^{\SPD}$ and  $\lambda_{\opt}^{\ThD}$ are indicated by horizontal lines up to the optimal numbers of multiplexed units $N_\opt$.
\label{fig:2}}
\end{figure*}

In our optimization procedure the objective function is the single-photon output probability $P_1$ while the variables of the optimization are the input mean photon numbers $\lambda_n$ in the multiplexed units.
The parameters of the optimization such as the number of generations, the populations, and the function tolerance are chosen appropriately to yield a reproducible and stable solution.

As it will be shown later, in the considered SPS the achievable single-photon probabilities $P_{1,N}^{S,\lambda_{n}}$ for any detection strategy $S$ are monotonically increasing functions of the number of multiplexed units $N$ and they saturate with the increase of the number of multiplexed units. Therefore, in order to determine a reasonable value of the optimal number of multiplexed units $N_\opt$ we first choose a reference value for the maximal single-photon probability  $P_{1,N_\text{ref}}^{S,\lambda_{n}}$ calculated at a high number of multiplexed units $N_\text{ref}$. We have found that for the choice of $N_\text{ref}=100$ the single-photon probability practically saturates, that is, it reaches its maximum in all our considered cases. Then we choose the value of $N_\opt$ to be equal to the smallest $N$ so that the difference between the corresponding single-photon probability $P_{1,N_\opt}^{S,\lambda_n}$ and the reference probability $P_{1,N_\text{ref}}^{S,\lambda_n}$  becomes less than $10^{-3}$. In the following, instead of $P_{1,N_\opt}^{S,\lambda_n}$, we will use the notation  $P_{1,\max}^{S,\lambda_n}$ because this value is the maximal single-photon probability that can be achieved by the given system using $N_\opt$ multiplexed units and it is practically equal to the saturated single-photon probability.

First we analyze the case when single-photon detection (SPD) is used for the heralding of the signal photons in the multiplexed units for all the considered ranges of the parameters $V_r$, $V_D$, and $V_b$.
Next we repeat the optimization procedure for the detection strategy $S=\{1,2\}$, that is, when the maximum accepted photon number is $J=2$.
We increase the value of $J$ until the maximal single-photon probability with the current detection strategy becomes less than the maximal single-photon probability achieved with the previous one. 
We also perform the optimization assuming threshold detection (ThD), that is, for $S=\mathbb{Z}^+$.
In addition, we determine the maximal single-photon probabilities $P_{1,\max}^{S,\lambda}$ that can be achieved in SPS with all the considered detection strategies assuming identical input mean photon numbers $\lambda_n=\lambda_{\opt}$ for each of the multiplexed units in order to clarify the advantage of optimizing the input mean photon numbers unit-wise.

In Fig.~\ref{fig:2}(a) we have plotted the achievable single-photon probabilities $P_{1,N}$ as a function of the number of multiplexed units $N$ for the SPD and ThD strategies assuming different and identical optimal input mean photon numbers $\lambda_{n,\opt}$ and $\lambda_{\opt}$, respectively, for each of the multiplexed units for the reflection efficiency $V_r=0.99$, the general transmission coefficient $V_b=0.98$, and the detector efficiency $V_D=0.9$.
From the figure one can deduce that the achievable single-photon probability $P_{1,N}$ saturates with the increasing number of multiplexed units $N$, as we already noted before.
The figure shows that regarding the achievable maximal single-photon probability $P_{1,N}$ SPD outperforms the ThD strategy for all considered values of the number of multiplexed units $N$.
As it can be expected, using different optimal input mean photon numbers $\lambda_{n,\opt}$ for each of the multiplexed units results in higher achievable single-photon probabilities $P_{1,N}$ compared to the case of using identical optimal input mean photon numbers $\lambda_\opt$ unit-wise for the considered detection strategies.
The enhancement that can be observed is significant only in the case of the ThD strategy, for SPD the increase is moderate.

\begin{figure*}
\centering
\includegraphics[width=\columnwidth]{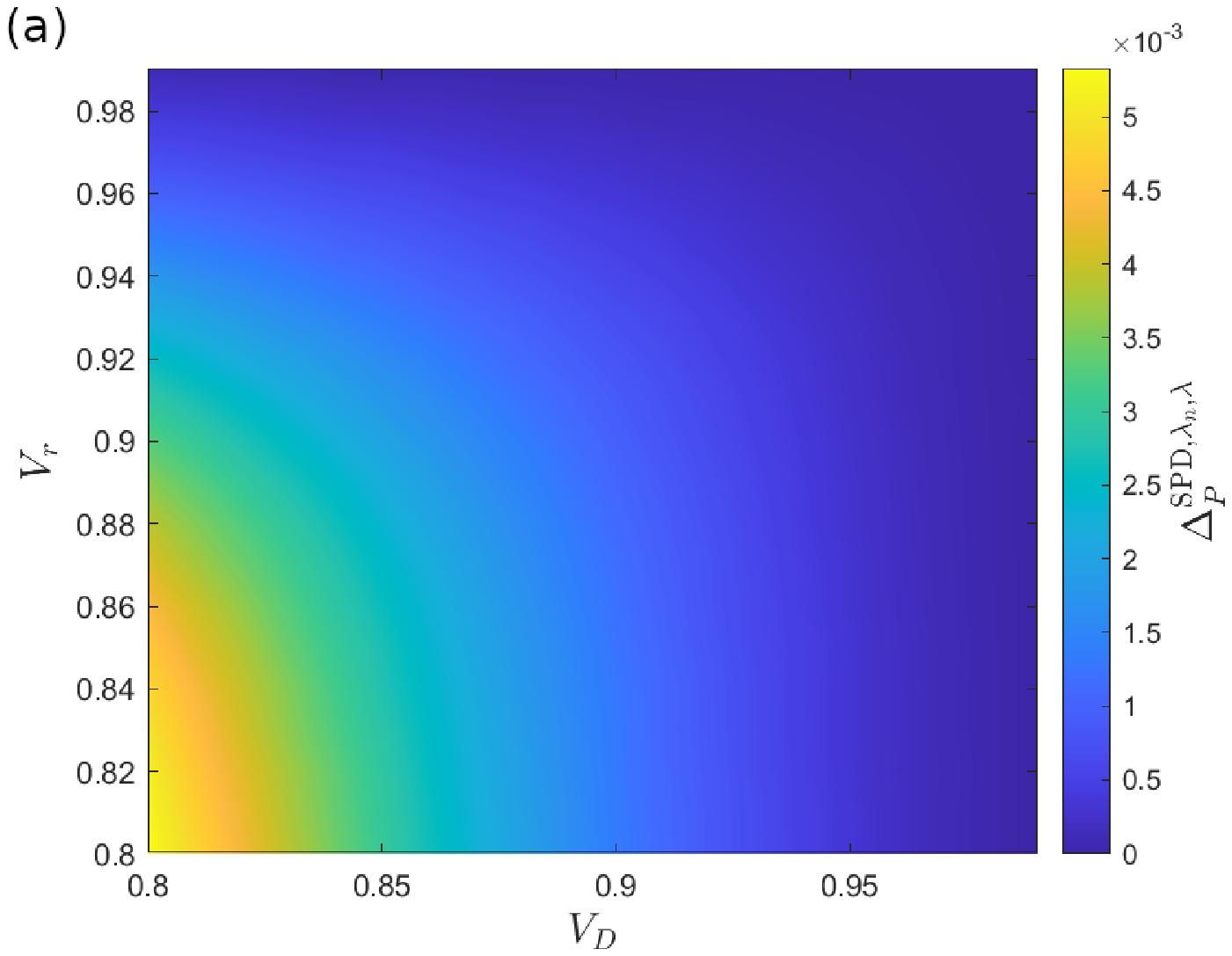}
\includegraphics[width=\columnwidth]{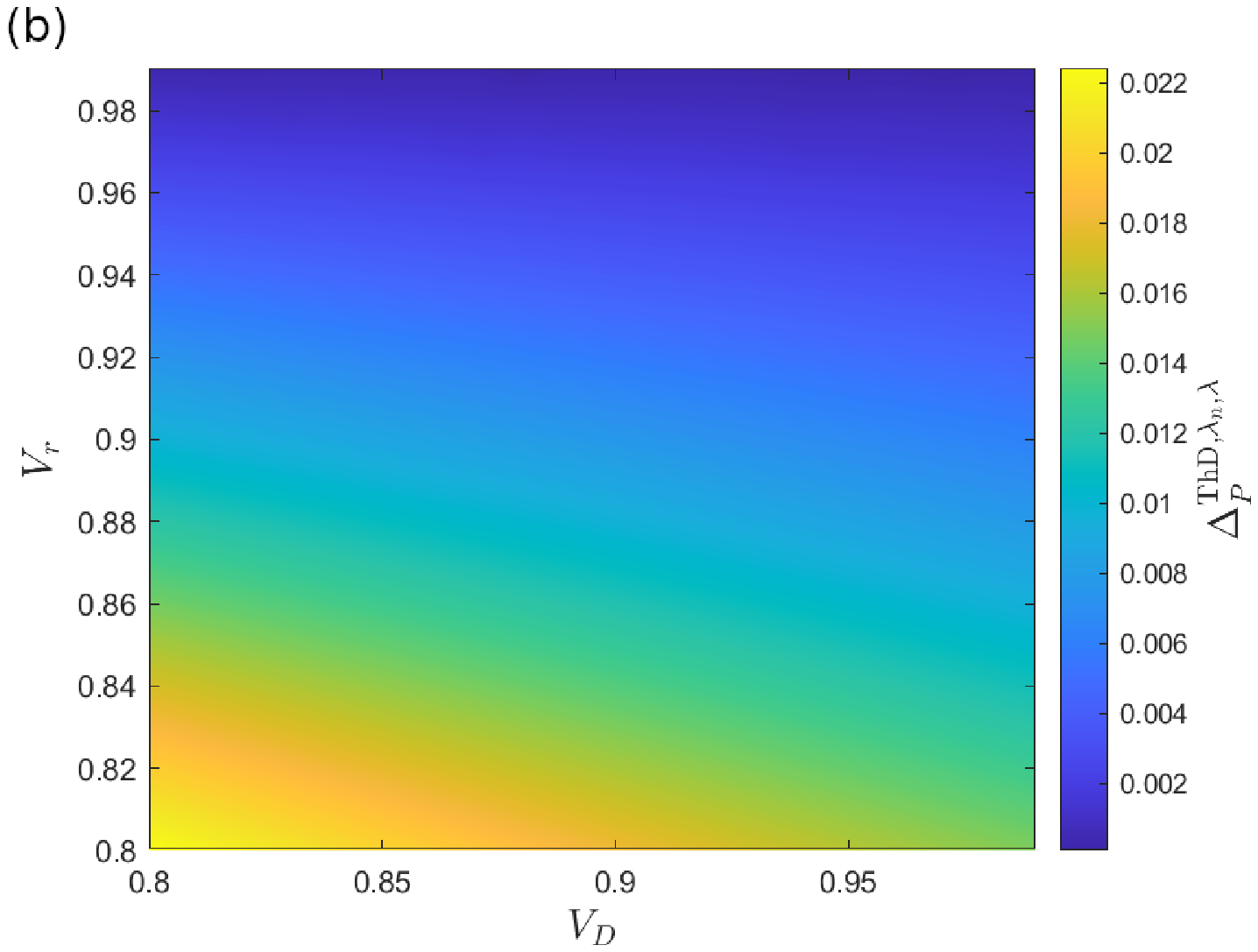}
\caption{(a) The difference $\Delta_P^{\SPD,\lambda_n,\lambda}=P_{1,\max}^{\SPD,{\lambda_{n}}}-P_{1,\max}^{\SPD,{\lambda}}$ between the maximal single-photon probabilities for the SPD strategy obtained by assuming different and identical optimal input mean photon numbers $\lambda_{n,\opt}$ and $\lambda_{\opt}$, respectively, in each of the multiplexed units, as a function of the detector efficiency $V_D$ and the reflection efficiency $V_r$ for the general transmission coefficient $V_b=0.98$.
(b) The corresponding difference $\Delta_P^{\ThD,\lambda_n,\lambda}=P_{1,\max}^{\ThD,\lambda_{n,}}-P_{1,\max}^{\ThD,\lambda}$ between the maximal single-photon probabilities for ThD as a function of the detector efficiency $V_D$ and the reflection efficiency $V_r$ for the general transmission coefficient $V_b=0.98$. \label{fig:3}}
\end{figure*}

The optimal input mean photon numbers $\lambda_{n,\opt}$ are presented in Fig.~\ref{fig:2}(b) as a function of the sequential number $n$ of the multiplexed units, for the considered detection strategies, up to the corresponding optimal numbers of multiplexed units.
The values of the optimal input mean photon numbers $\lambda_{\opt}^{\SPD}$ and $\lambda_{\opt}^{\ThD}$ under the assumption of being identical for all the units are indicated by continuous and dashed lines, respectively, up to the optimal numbers of multiplexed units.
From the data plotted in Fig.~\ref{fig:2}(b) one can conclude that with increasing sequential number $n$, i.e., increasing losses the optimal input mean photon numbers $\lambda_{n,\opt}$ also increase.
For the first few multiplexed units the optimal input mean photon numbers $\lambda_{n,\opt}$ remain below the value $\lambda_{\opt}$ (i.e. the optimal one assuming all $\lambda$'s are identical). For higher sequential numbers $n$, however, there are higher input mean photon numbers that compensate for higher losses, in case of all the considered strategies.

In Fig.~\ref{fig:3}(a) we have plotted the difference $\Delta_P^{\SPD,\lambda_n,\lambda}=P_{1,\max}^{\SPD,\lambda_{n}}-P_{1,\max}^{\SPD,\lambda}$ between the maximal single-photon probabilities for the SPD strategy obtained by assuming both different and identical optimal input mean photon numbers $\lambda_{n,\opt}$ and $\lambda_{\opt}$, respectively, in each of the multiplexed units, as a function of the detector efficiency $V_D$ and the reflection efficiency $V_r$ for the general transmission coefficient $V_b=0.98$.
With the decrease of the detector efficiency $V_D$ and the reflection efficiency $V_r$ the difference $\Delta_P^{\SPD,\lambda_n,\lambda}$ increases.
This effect can be intuitively explained as follows.
Even in the case of the SPD strategy the decrease of the detector efficiency $V_D$ leads to an increase in the probability of more than one photons entering the multiplexer.
This can be advantageous in the case of higher losses, that is, multiplexed units with higher sequential numbers $n$ and in systems with lower reflection efficiencies $V_r$.
This advantage can be fully exploited by optimizing the input mean photon numbers $\lambda_n$ for each of the multiplexed units separately.
In the considered parameter ranges of the reflection and detector efficiencies $V_r$ and $V_D$, respectively, the largest difference was found to be $\Delta_P^{\SPD,\lambda_n,\lambda}\simeq 0.006$ at $V_r=0.8$ and $V_D=0.8$, that is, for the lowest values of the reflection and detector efficiencies of the considered ranges.

In Fig.~\ref{fig:3}(b) we have plotted the corresponding difference $\Delta_P^{\ThD,\lambda_n,\lambda}=P_{1,\max}^{\ThD,{\lambda_{n}}}-P_{1,\max}^{\ThD,{\lambda}}$ of the maximal single-photon probabilities for the ThD strategy.
The difference $\Delta_P^{\ThD,\lambda_n,\lambda}$, that is, the enhancement in the maximal single-photon probabilities $P_{1,\max}^{\ThD}$ using different optimal input mean photon numbers $\lambda_{n,\opt}$ instead of identical ones $\lambda_{\opt}$ for each of the multiplexed units is always higher than the corresponding difference $\Delta_P^{\SPD,\lambda_n,\lambda}$ for SPD strategy for the whole parameter ranges of the reflection and detection efficiencies $V_r$ and $V_D$, respectively.
\begin{figure*}[tb]
    \centering
    \includegraphics[width=0.99\columnwidth]{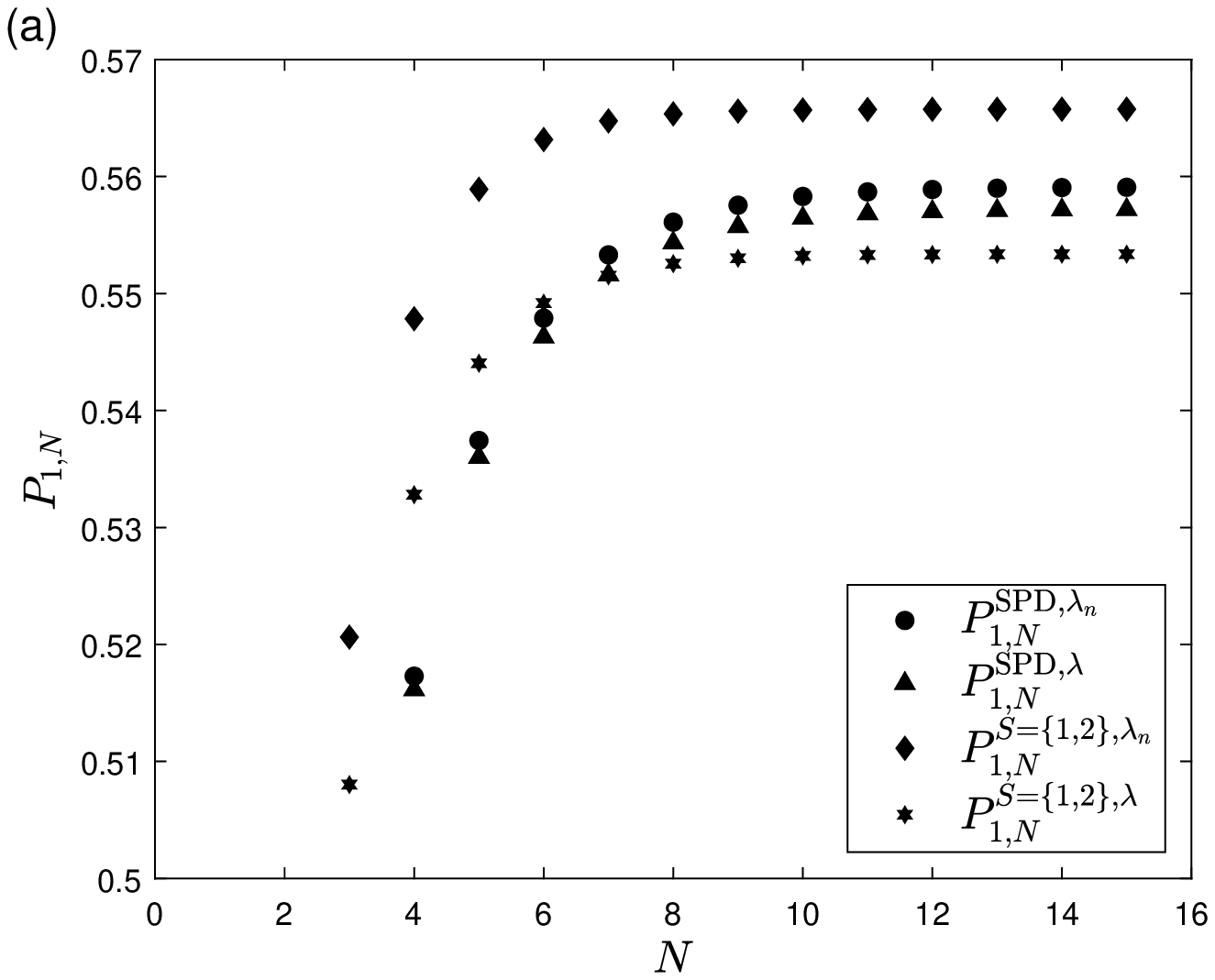}\hfill
    \includegraphics[width=0.99\columnwidth]{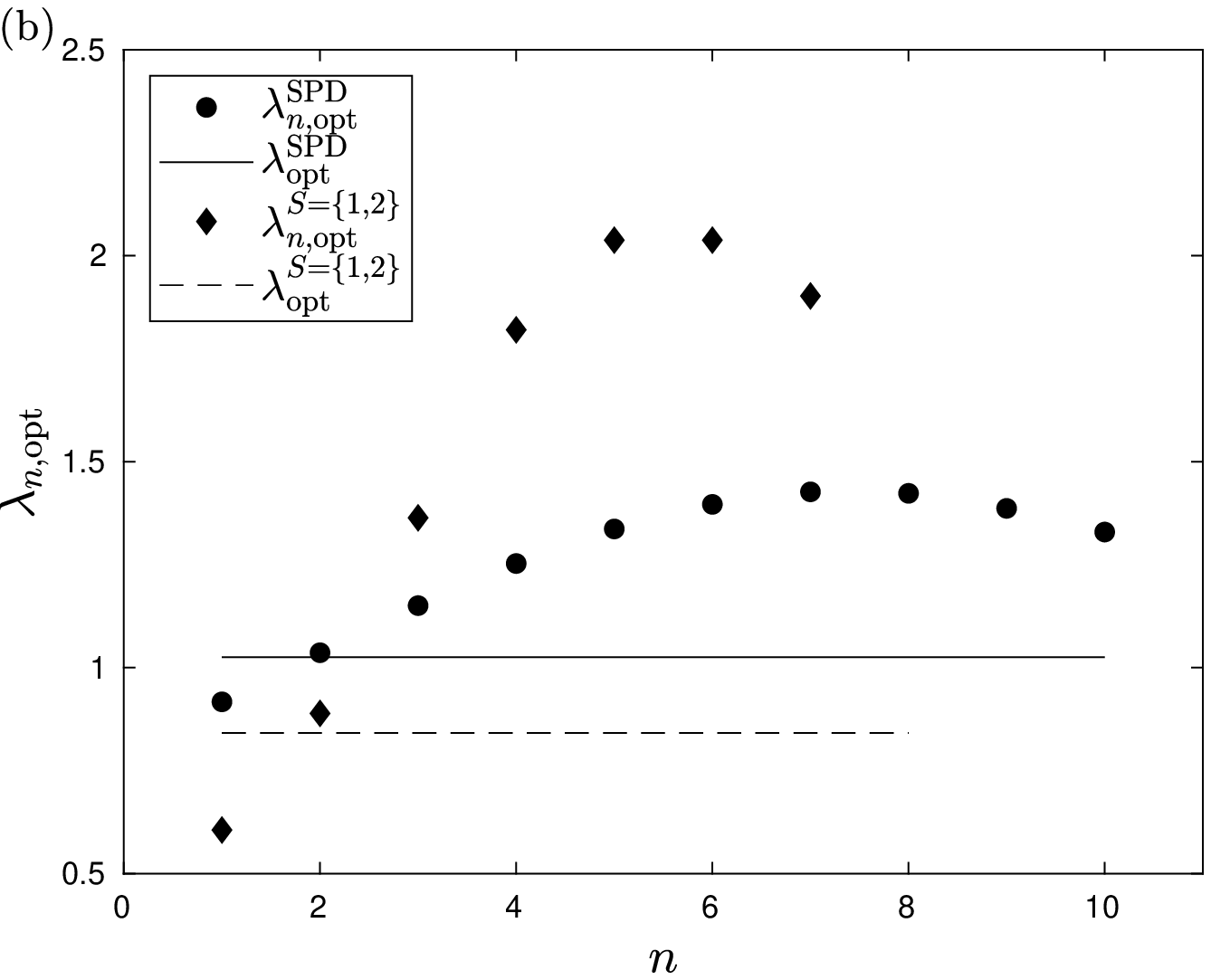}
\caption{(a) The achievable single-photon probabilities $P_{1,N}$ as a function of the number of multiplexed units $N$ for the SPD and $S=\{1,2\}$ detection strategies assuming different and identical optimal input mean photon numbers $\lambda_{n, \opt}$ and $\lambda_{\opt}$, respectively, for each of the multiplexed units for the reflection efficiency $V_r=0.8$, the general transmission coefficient $V_b=0.8$, and the detector efficiency $V_D=0.85$.
(b) Optimal input mean photon numbers $\lambda_{n,\opt}$ as a function of the sequential number $n$ of the multiplexed units for the considered detection strategies up to the corresponding optimal numbers of multiplexed units and for the parameters of Fig.~\ref{fig:4}(a). The values of the identical optimal input mean photon numbers $\lambda_{\opt}^{\SPD}$ and $\lambda_{\opt}^{S=\{1,2\}}$ are indicated by horizontal lines up to the optimal numbers of multiplexed units.
\label{fig:4}}
\end{figure*}
In the range $0.8\leq V_D\leq 0.98$ and $0.8\leq V_r\leq 0.86$ the enhancement is above 1\%, the highest enhancement is $\Delta_P^{\ThD,\lambda_n,\lambda}=0.0225$ at $V_r=0.8$ and $V_D=0.8$.

Let us now address the question of the optimal choice of the detection strategy.
As an example illustrating the role of the strategy, in Fig.~\ref{fig:4}(a) we have plotted the achievable single-photon probabilities $P_{1,N}$ as a function of the number of multiplexed units $N$ for the SPD and $S=\{1,2\}$ detection strategies assuming different and identical optimal input mean photon numbers $\lambda_{n,\opt}$ and $\lambda_{\opt}$, respectively, for each of the multiplexed units. The reflection efficiency is $V_r=0.8$, the general transmission coefficient is $V_b=0.8$, and the detector efficiency is $V_D=0.85$ here, which are smaller than those used in Fig.~\ref{fig:2}.
Obviously, the achievable single-photon probabilities $P_{1,N}$ obtained with this set of parameters are also smaller than the values presented in Fig.~\ref{fig:2}.
In the presented case the values of the achievable single-photon probabilities $P_{1,N}^{S=\{1,2\},\lambda_{n}}$ for the detection strategy $S=\{1,2\}$ obtained at different optimal input mean photon numbers $\lambda_{n,\opt}$ are the highest.
From Fig.~\ref{fig:4}(a) one can also deduce that in the case of the $S=\{1,2\}$ detection strategy larger enhancements can be achieved in the maximal single-photon probabilities by the unit-wise optimization of the input mean photon numbers $\lambda_n$ than the enhancements achievable for the SPD strategy.

Interestingly, for the number of multiplexed units $N<7$ the single-photon probabilities $P_{1,N}^{S=\{1,2\},\lambda}$ achievable by applying the $S=\{1,2\}$ detection strategy and using identical optimal input mean photon numbers $\lambda_\opt$ in each of the multiplexed units is higher than the corresponding probabilities $P_{1,N}^{\SPD,\lambda}$ or $P_{1,N}^{\SPD,\lambda_n}$ achievable by applying the SPD strategy with either identical or different input mean photon numbers, respectively.
However, above this value ($N\ge7$) the relationship between these quantities is reversed, that is,  $P_{1,N}^{\SPD,\lambda}> P_{1,N}^{S=\{1,2\},\lambda}$ and $P_{1,N}^{\SPD,\lambda_n}> P_{1,N}^{S=\{1,2\},\lambda}$.
In brief, given a number of multiplexed units, the choice of the optimal strategy can depend on whether the input photon numbers are identical or different.

In Fig.~\ref{fig:4}(b) we present the optimal input mean photon numbers $\lambda_{n,\opt}$ as a function of the sequential number $n$ of the multiplexed units for the considered detection strategies up to the corresponding optimal numbers of multiplexed units and for the parameters of Fig.~\ref{fig:4}(a). The values of the identical optimal input mean photon numbers $\lambda_{\opt}^{\SPD}$ and $\lambda_{\opt}^{S=\{1,2\}}$ are indicated by horizontal lines up to the optimal numbers of multiplexed units.
An interesting feature of the values of the input mean photon numbers $\lambda_{n,\opt}$ is that, in contrast to the similar curve presented in Fig.~\ref{fig:2}(b), in this case a well-defined peak (maximum) can be observed at $n=6$ for the strategy $S=\{1,2\}$, and at $n=8$ for SPD.
A plausible explanation for this behavior is that for high values of the sequential numbers $n$ the total transmission coefficient $V_n$ of the $n$th multiplexed unit is so low that it is more advantageous to decrease the probability of the detection events for units with high sequential numbers $n$, thus suppressing their participation in the multiplexing process.

\begin{figure}[tb]
    \centering
    \includegraphics[width=\columnwidth]{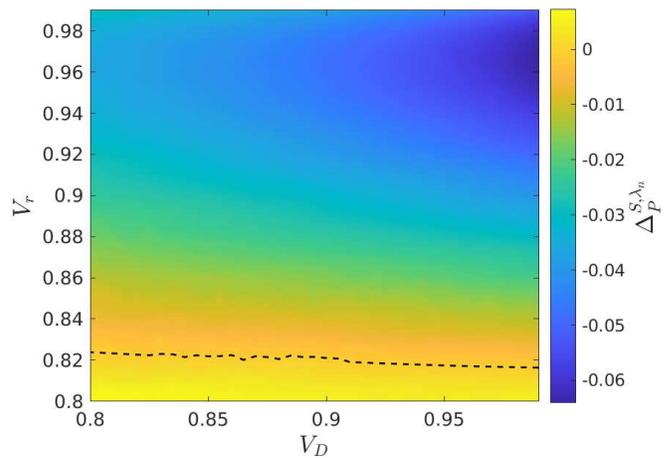}
\caption{The difference  $\Delta_P^{S,\lambda_n}=P_{1,\max}^{S=\{1,2\},\lambda_n}-P_{1,\max}^{\SPD,\lambda_n}$ between the maximal single-photon probabilities obtained by assuming different optimal input mean photon numbers $\lambda_{n,\opt}$ in each of the multiplexed units with two types of detection strategies: for those defined by the set $S=\{1,2\}$ and for SPD, as a function of the detector efficiency $V_D$ and the reflection efficiency $V_r$ for 
the general transmission coefficient $V_b=0.8$. Below the dashed line the detection strategy $S=\{1,2\}$ outperforms the SPD strategy.
\label{fig:5}}
\end{figure}

\begin{table*}[!ht]
\caption{Maximal single-photon probabilities $P^{\text{SPD},\lambda_n}_{1,\max}$ for the SPD strategy obtained at different optimal input mean photon numbers $\lambda_{n,\opt}^{\text{SPD}}$ and the optimal number of multiplexed units $N_\opt^{\text{SPD},\lambda_n}$ for various values of the general transmission coefficient $V_b$, the detector efficiency $V_D$, and the reflection efficiency $V_r$.
The optimal number of multiplexed units $N_{\opt}^{\text{SPD},\lambda}$ required to achieve maximal single-photon probabilities $P^{\text{SPD},\lambda}_{1,\max}$ obtained at identical optimal input mean photon numbers $\lambda_{\opt}^{\text{SPD}}$, and the optimal identical input mean photon numbers $\lambda_\opt^{\text{SPD}}$ at which these probabilities can be achieved are also displayed.
\label{tab:3}}
\begin{ruledtabular}
\begin{tabular}{c|c|cccc|cccc|cccc}
      &&\multicolumn{4}{c|}{$V_b=0.8$}&\multicolumn{4}{c|}{$V_b=0.9$}
      &\multicolumn{4}{c}{$V_b=0.98$}\\\hline
\rule{0pt}{2.6ex} $V_r$& $V_D$ &
\multicolumn{1}{c}{$P_{1,\max}^{\text{SPD},\lambda_n}$}&$N_\opt^{\text{SPD},\lambda_n}$&$N_\opt^{\text{SPD},\lambda}$&$\lambda_{\opt}^{\text{SPD}}$&
\multicolumn{1}{c}{$P_{1,\max}^{\text{SPD},\lambda_n}$}&$N_\opt^{\text{SPD},\lambda_n}$&$N_\opt^{\text{SPD},\lambda}$&$\lambda_{\opt}^{\text{SPD}}$&
\multicolumn{1}{c}{$P_{1,\max}^{\text{SPD},\lambda_n}$}&$N_\opt^{\text{SPD},\lambda_n}$&$N_\opt^{\text{SPD},\lambda}$&$\lambda_{\opt}^{\text{SPD}}$\\\hline
& 0.80  & 0.622 & 13 & 13 & 0.859 & 0.681 & 13 & 14 & 0.771 & 0.728 & 13 & 14 & 0.716 \\
& 0.85 & 0.634 & 12 & 13 & 0.891 & 0.698 & 13 & 13 & 0.815 & 0.749 & 13 & 13 & 0.765 \\
& 0.90  & 0.646  & 12 & 12 & 0.929 & 0.716  & 12 & 13 & 0.868 & 0.771 & 13 & 13 & 0.826 \\
0.90  & 0.92 & 0.651  & 12 & 12 & 0.943 & 0.723  & 12 & 13 & 0.892 & 0.781  & 13 & 13 & 0.856 \\
& 0.94 & 0.656  & 12 & 12 & 0.958 & 0.731  & 12 & 12 & 0.919 & 0.790   & 12 & 13 & 0.888 \\
& 0.96 & 0.661  & 12 & 12 & 0.973 & 0.739  & 12 & 12 & 0.945 & 0.801  & 12 & 13 & 0.923 \\
& 0.98 & 0.666  & 12 & 12 & 0.987 & 0.747  & 12 & 12 & 0.973 & 0.811  & 12 & 12 & 0.962 \\\hline
& 0.80  & 0.669  & 14 & 16 & 0.671 & 0.737 & 15 & 17 & 0.592 & 0.790  & 16 & 18 & 0.543 \\
& 0.85 & 0.682  & 14 & 15 & 0.722 & 0.754  & 15 & 16 & 0.643 & 0.810   & 15 & 17 & 0.593 \\
& 0.90  & 0.695  & 14 & 14 & 0.793 & 0.771  & 14 & 15 & 0.719 & 0.832  & 15 & 16 & 0.670  \\
0.95 & 0.92 & 0.700    & 14 & 14 & 0.825 & 0.779  & 14 & 15 & 0.757 & 0.841  & 14 & 15 & 0.714 \\
& 0.94 & 0.706  & 14 & 14 & 0.862 & 0.787  & 14 & 14 & 0.807 & 0.851  & 14 & 15 & 0.764 \\
& 0.96 & 0.712  & 13 & 14 & 0.904 & 0.796  & 14 & 14 & 0.861 & 0.863  & 14 & 14 & 0.830  \\
& 0.98 & 0.718  & 13 & 13 & 0.952 & 0.806  & 14 & 14 & 0.926 & 0.875  & 14 & 14 & 0.907 \\\hline
& 0.80  & 0.732  & 23 & 28 & 0.344 & 0.814  & 25 & 32 & 0.294 & 0.880   & 28 & 35 & 0.267 \\
& 0.85 & 0.739  & 20 & 25 & 0.383 & 0.824  & 23 & 28 & 0.331 & 0.892  & 25 & 31 & 0.298 \\
& 0.90  & 0.748  & 18 & 21 & 0.455 & 0.836  & 20 & 24 & 0.391 & 0.905  & 21 & 26 & 0.355 \\
0.99 & 0.92 & 0.752  & 18 & 20 & 0.494 & 0.841  & 19 & 22 & 0.431 & 0.911  & 20 & 24 & 0.391 \\
& 0.94 & 0.756  & 17 & 18 & 0.558 & 0.846  & 18 & 20 & 0.486 & 0.918  & 19 & 22 & 0.440  \\
& 0.96 & 0.761  & 16 & 17 & 0.637 & 0.853  & 17 & 18 & 0.570  & 0.925  & 17 & 19 & 0.526 \\
& 0.98 & 0.767  & 15 & 15 & 0.781 & 0.860   & 16 & 16 & 0.715 & 0.935  & 16 & 17 & 0.667\\\hline
\end{tabular}
\end{ruledtabular}
\end{table*}

After analyzing in detail a case where the detection strategy $S=\{1,2\}$ outperforms the SPD strategy, in Fig.~\ref{fig:5} we show the difference  $\Delta_P^{S,\lambda_n}=P_{1,\max}^{S=\{1,2\},\lambda_n}-P_{1,\max}^{\SPD,\lambda_n}$ between the maximal single-photon probabilities obtained by assuming different optimal input mean photon numbers $\lambda_{n,\opt}$ in each of the multiplexed units with the given detection strategies as a function of the detector efficiency $V_D$ and the reflection efficiency $V_r$ for the general transmission coefficient $V_b=0.8$. Below the dashed line in the figure the detection strategy $S=\{1,2\}$ outperforms the SPD strategy.
At the reflection efficiency $V_r=0.8$ the difference is $\Delta_P^{S,\lambda_n}>0.005$ for the whole range of $V_D$, and the highest difference is $\Delta_P^{S,\lambda_n}=0.007$ at $V_D=0.8$.
When analyzing the effect of the general transmission coefficient $V_b$ we have found that as we replot the figure with an increased value of $V_b$, the dashed line in the figure gets shifted towards the horizontal axis, that is, towards smaller values of $V_r$. As we reach the value $V_b=0.837$ there are no more points left in the figure for which $\Delta_P^{S,\lambda_n}\ge0$.
This implies that for values of the general transmission coefficient $V_b>0.837$ SPD is the optimal detection strategy.
We note that over the considered ranges of the parameters $V_r$, $V_D$, and $V_b$ there are no cases where the detection strategy defined by the set $S=\{1,2,3\}$ would lead to higher single-photon probabilities than the ones that can be obtained by using either the SPD or the  $S=\{1,2\}$ detection strategies. 

Let us now assess the performance of our scheme in more detail. 
In Table~\ref{tab:3} we have presented the maximal single-photon probabilities $P^{\SPD,\lambda_n}_{1,\max}$ for the SPD strategy obtained at different optimal input mean photon numbers $\lambda_{n,\opt}^{\SPD}$, and the required number of multiplexed units $N_\opt^{\SPD,\lambda_n}$ for various values of the general transmission coefficient $V_b$, the detector efficiency $V_D$, and the reflection efficiency $V_r$.
We have also displayed the optimal number of multiplexed units $N_{\opt}^{\SPD,\lambda}$ required to achieve maximal single-photon probabilities $P^{\SPD,\lambda}_{1,\max}$ obtained at identical optimal input mean photon numbers $\lambda_{\opt}^{\SPD}$, and the optimal input mean photon numbers $\lambda_\opt^{\SPD}$ at which these probabilities can be achieved.

From data of Table~\ref{tab:3} one can deduce that the highest single-photon probability $P^{\SPD,\lambda_n}_{1,\max}$ that can be achieved by SPS based on asymmetric spatial multiplexing using SPD strategy obtained at different optimal input mean photon numbers $\lambda_{n,\opt}^{\SPD}$ is $P^{\SPD,\lambda_n}_{1,\max}=0.935$.
This single-photon probability is, up to our knowledge, the highest one reported in the literature that, in principle, can be produced by multiplexed SPS using state-of-the-art experimental parameters.
It can be achieved at the best experimentally realizable parameter values that are the reflection efficiency $V_r=0.99$, the general transmission coefficient $V_b=0.98$, and the detector efficiency $V_D=0.98$.
\begin{figure*}[tb]
    \centering
    \includegraphics[width=\columnwidth]{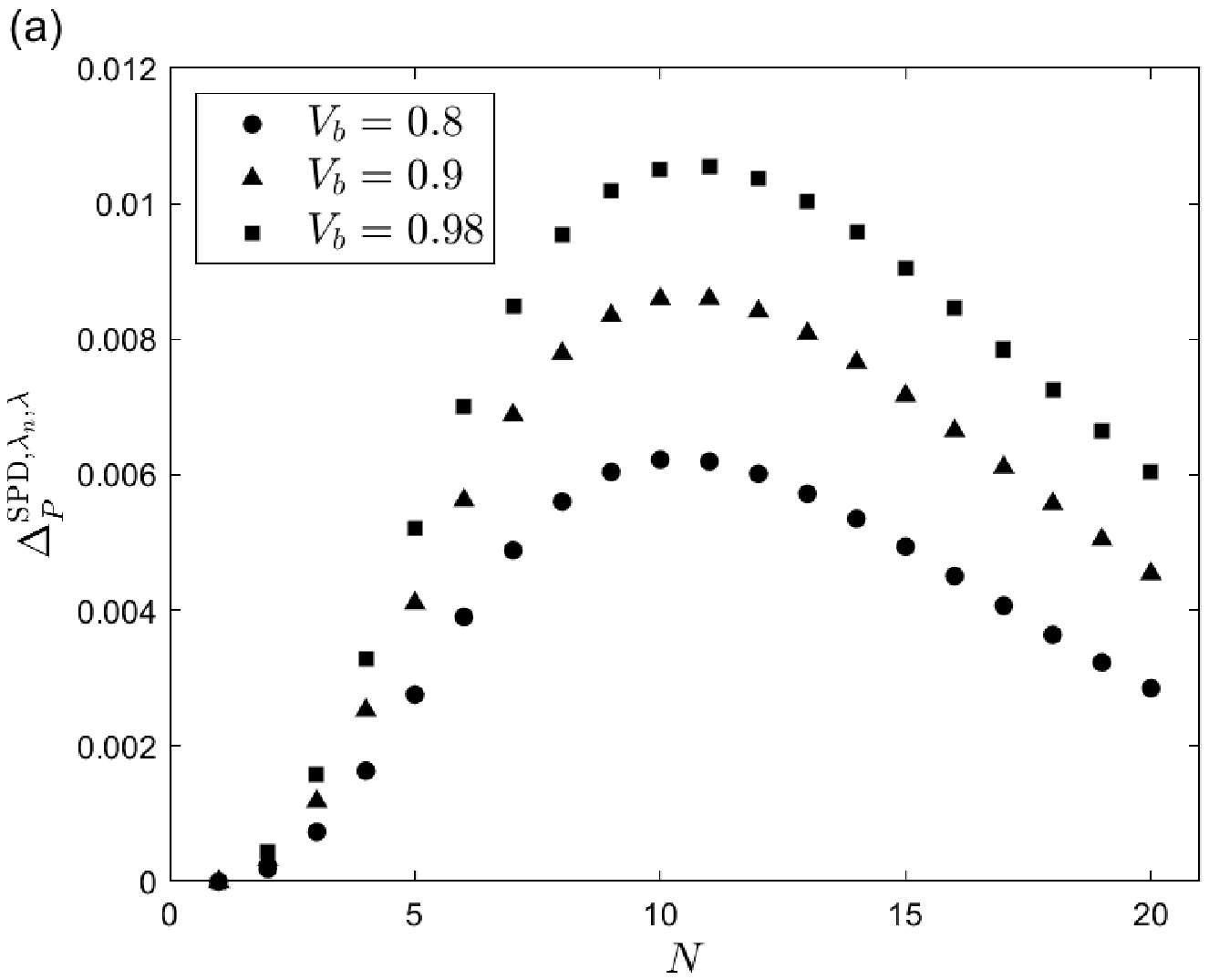}
    \includegraphics[width=\columnwidth]{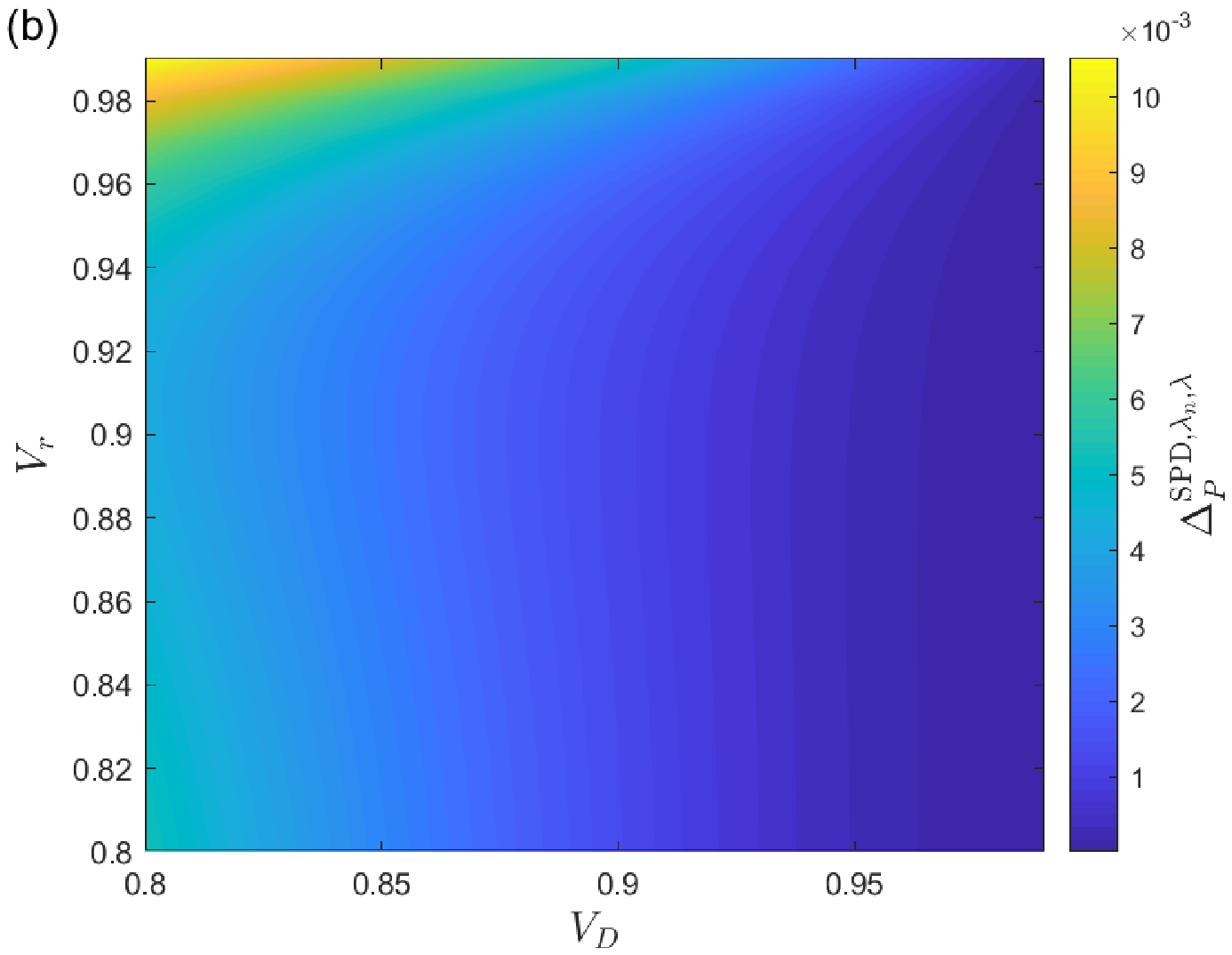}
\caption{\label{fig:6}(a) The difference $\Delta_P^{\SPD,\lambda_n,\lambda}=P_{1,N}^{\SPD,\lambda_n}-P_{1,N}^{\SPD,\lambda}$ between the achievable single-photon probabilities for SPD obtained by assuming different and identical optimal input mean photon numbers $\lambda_{n,\opt}$ and $\lambda_{\opt}$, respectively, in each of the multiplexed units, as a function of the number of multiplexed units $N$ for the reflection efficiency $V_r=0.99$ and the detector efficiency $V_D=0.8$ in the case of three different values of the general transmission coefficients $V_b$. (b) The same quantity $\Delta_P^{\SPD,\lambda_n,\lambda}$ as a function of the reflection efficiency $V_r$ and the detector efficiency $V_D$ for the general transmission coefficient $V_b=0.98$, and the number of multiplexed units $N=11$.}
\end{figure*}

Table \ref{tab:3} shows that the optimal numbers of multiplexed units $N_\opt^{\SPD,\lambda_n}$ and $N_\opt^{\SPD,\lambda}$ increase by increasing the values of the general transmission coefficient $V_b$ or the reflection efficiency $V_r$, while an increase in the detector efficiency $V_D$ leads to a decrease in them.
This property is not unexpected because in the case of smaller losses, that is, for higher transmissions larger optimal multiplexed system sizes can be used to increase the maximal single-photon probability.
However, if the detector efficiency $V_D$ increases, that is, the probability that the detectors select the single-photon events correctly increases, then the detectors are sufficient themselves to exclude the multi-photon events, thus suppressing these events via decreasing the intensity and increasing the system size becomes less crucial.

Table~\ref{tab:3} also shows that the optimal numbers of multiplexed units $N_{\opt}^{\SPD,\lambda_n}$ obtained at different optimal input mean photon numbers $\lambda_{n,\opt}$ are less than or equal to the corresponding quantity $N_{\opt}^{\SPD,\lambda}$ obtained at identical optimal input mean photon numbers $\lambda_{\opt}$, that is,  $N_{\opt}^{\SPD,\lambda_n}\le N_{\opt}^{\SPD,\lambda}$.
This observation is in accordance with the findings related to Figs.~\ref{fig:2}(b) and \ref{fig:4}(b).
Hence, the unit-wise optimization of the input mean photon numbers can result in the decrease of the optimal system size needed to maximize the single-photon probability. This can be an advantage in an experimental realization of the system.
Larger differences between the optimal system sizes occur for smaller values of the detector efficiency $V_D$ and for larger values of the reflection efficiency $V_r$.
The potential decrease of the system size justifies the application of this kind of optimization, in spite of not resulting in dramatic improvement in the achievable maximal single-photon probability for the SPD strategy.

Concerning the optimal identical input mean photon number $\lambda_{\opt}^{\SPD}$, with the increase of the value of either the reflection efficiency $V_r$ or the general transmission coefficient $V_b$ the value of $\lambda_{\opt}^{\SPD}$ decreases while its value increases with increasing the detector efficiency $V_D$.
Intuitively, in the case of fewer multiplexed units $N_{\opt}^{\SPD,\lambda}$ increasing the optimal input mean photon number $\lambda_{\opt}^{\SPD}$ guarantees that at least one heralding event occurs in the whole multiplexed system.

Let us also address the advantage that can be achieved by the unit-wise optimization of the input mean photon numbers $\lambda_n$ in the case of suboptimal numbers of multiplexed units $N$. This is the typical situation in the experiments realized so far \cite{Ma2010, KiyoharaOE2016, Meany2014, Collins2013, Francis2016}.
To address this question, in Fig.~\ref{fig:6}(a) we have plotted the difference $\Delta_P^{\SPD,\lambda_n,\lambda}=P_{1,N}^{\SPD,\lambda_n}-P_{1,N}^{\SPD,\lambda}$ between the achievable single-photon probabilities for SPD obtained by assuming different and identical optimal input mean photon numbers $\lambda_{n,\opt}$ and $\lambda_{\opt}$, respectively, in each of the multiplexed units, as a function of the number of multiplexed units $N$ for the reflection efficiency $V_r=0.99$ and the detector efficiency $V_D=0.8$ in the case of three different general transmission coefficients $V_b$.
The figure shows that for these parameters the enhancements $\Delta_P^{\SPD,\lambda_n,\lambda}$ depending on the number of multiplexed units $N$ have maxima that increase with increasing the values of the general transmission coefficient  $V_b$ and the locations of the peaks for $V_b=0.8$ and $V_b=0.9$ are at $N=10$ while for $V_b=0.98$ it is at $N=11$.
At this number of multiplexed units the achievable single-photon probability is $P_{1,N=11}^{\SPD,\lambda_n}=0.846$. For $V_b=0.98$ the enhancement is $\Delta_P^{\SPD,\lambda_n,\lambda}\ge0.01$ for the range $9\le N\le13$.
Hence, though the enhancement in the maximal single-photon probabilities $P_{1,\max}^{\SPD}$ due to the unit-wise optimization of the input mean photon numbers is generally moderate (c.f.\ Fig.~\ref{fig:3}(a)), for the given set of suboptimal system sizes an observable enhancement can be obtained in the achievable single-photon probabilities at the considered parameters of the setup.

Figure~\ref{fig:6}(b) presents the same difference $\Delta_P^{\SPD,\lambda_n,\lambda}$ as a function of the reflection efficiency $V_r$ and the detector efficiency $V_D$ for the general transmission coefficient $V_b=0.98$ and the number of multiplexed units $N=11$.
The figure shows that the highest differences $\Delta_P^{\SPD,\lambda_n,\lambda}$ in the achievable single-photon probabilities can be achieved for high values of the reflection efficiency $V_r$ and low values of the detection efficiency $V_D$.
Thus the most notable improvement is achieved in the case of better routers.

\begin{figure}[!t]
    \centering
    \includegraphics[width=\columnwidth]{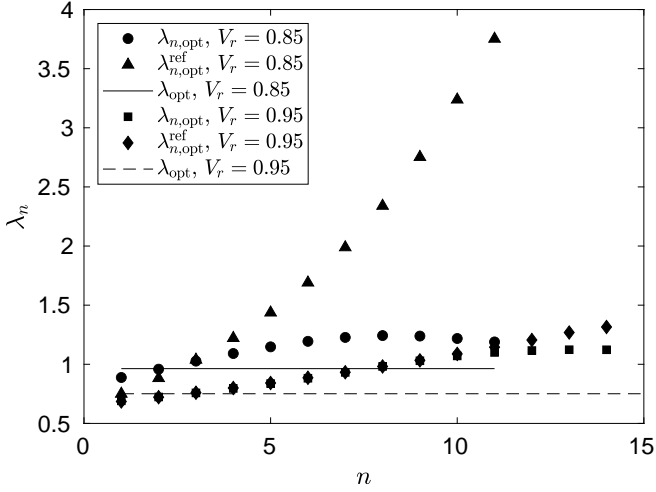}
    \caption{Different optimal input mean photon numbers $\lambda_{n,\opt}$ obtained by our optimization procedure, and  $\lambda_{n,\opt}^{\text{ref}}$ obtained by using  Eq.~\eqref{formula:maz} as a function of the sequential number $n$ of the multiplexed units for the SPD strategy up to the corresponding optimal numbers of multiplexed units
    for two values of the reflection efficiency $V_r=0.85$ and $V_r=0.95$, for the detector efficiency $V_D=0.9$ and the general transmission coefficient $V_b=0.85$. The values of the identical optimal input mean photon numbers $\lambda_{\opt}$ are indicated by horizontal lines up to the optimal numbers of multiplexed units.\label{fig:7}}
\end{figure}

\begin{figure}
    \centering
\includegraphics[width=\columnwidth]{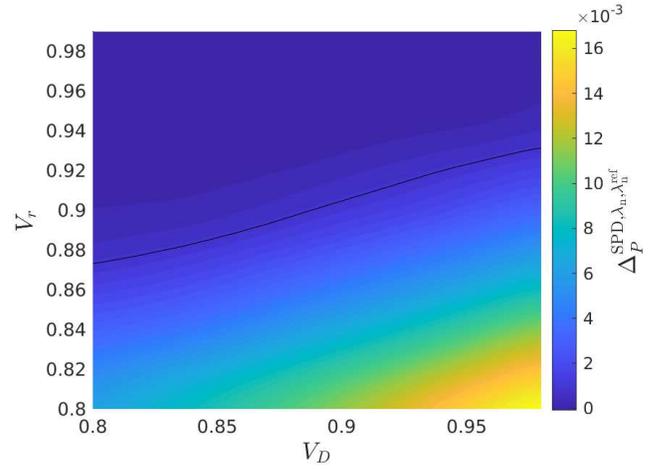}
    \caption{The difference $\Delta_P^{\SPD,\lambda_n,\lambda_n^{\text{ref}}}=P_{1,\max}^{\SPD,\lambda_n}-P_{1,\max}^{\SPD,\lambda_n^{\text{ref}}}$ between the maximal single-photon probabilities for SPD obtained by assuming different input mean photon numbers $\lambda_{n,\opt}$ and input mean photon numbers $\lambda_{n,\opt}^{\text{ref}}$ of Eq.~\eqref{formula:maz}, respectively, in each of the multiplexed units, as a function of the reflection efficiency $V_r$ and the detector efficiency $V_D$ for the general transmission coefficient $V_b=0.85$. Below the continuous black line the difference $\Delta_P^{\SPD,\lambda_n,\lambda_n^{\text{ref}}}$ is higher than $10^{-3}$.\label{fig:8}}
\end{figure}

Now let us investigate the relation of our results to the direct application of the proposal in Ref.~\cite{MazzarellaPRA2013} to our scheme. In our notation the functional dependence for the different input mean photon numbers proposed in that paper can be expressed as
\begin{equation}
    \lambda_{n}^{\text{ref}}=\frac{\lambda}{V_n}.\label{formula:maz}
\end{equation}
In this formula the subsequent input mean photon numbers are scaled up by the different total transmission coefficients $V_n$ corresponding to the different multiplexed units. The optimization of the input mean photon numbers $\lambda_n^{\rm ref}$ can be performed by substituting this formula into Eq.~\eqref{general_formula} and then applying a standard optimization procedure for the parameter $\lambda$.
In Fig.~\ref{fig:7} we present the different optimal input mean photon numbers $\lambda_{n,\opt}$ obtained via our optimization procedure, and $\lambda_{n,\opt}^{\text{ref}}$ obtained by using  Eq.~\eqref{formula:maz} as a function of the sequential number $n$ of the multiplexed units for the SPD strategy up to the corresponding optimal numbers of multiplexed units. The data are calculated for two values of the reflection efficiency $V_r=0.85$ and $V_r=0.95$, for the detector efficiency $V_D=0.9$ and the general transmission coefficient $V_b=0.85$.
The values of the identical optimal input mean photon numbers $\lambda_{\opt}$ are indicated by horizontal lines up to the optimal numbers of multiplexed units.
The figure shows that for smaller values of the reflection efficiency $V_r$ the different optimal input mean photon numbers $\lambda_{n,\opt}$ and $\lambda_{n,\opt}^{\text{ref}}$ obtained from our full statistical description and by using Eq.~\eqref{formula:maz}, respectively, are significantly different. For higher values of $V_r$ considerable differences can be observed only for higher sequential numbers $n$.

\begin{figure}[!t]
    \centering
    \includegraphics[width=\columnwidth]{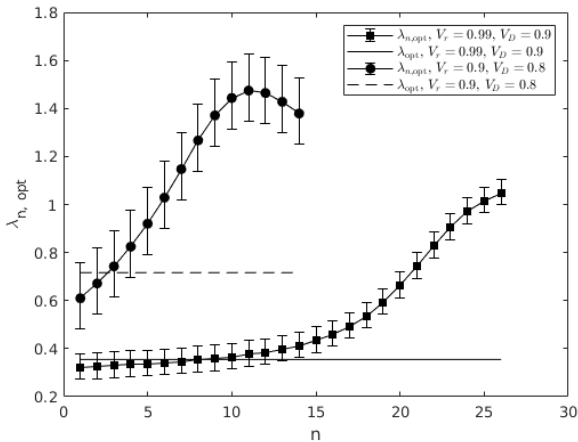}
\caption{The optimal input mean photon numbers $\lambda_{n,\opt}$ as a function of the sequential number $n$ for the SPD strategy for two pairs of the reflection and detector efficiencies $V_r=0.99$, $V_D=0.9$, and $V_r=0.9$, $V_D=0.8$, and for the general transmission coefficient $V_b=0.98$. The error bars show the amount of allowed uniform deviation from the optimal values of $\lambda_n$ so that the resulting single-photon probability is still higher than the one achievable by using identical optimal mean photon numbers $\lambda_{\opt}$. The values of the identical optimal input mean photon numbers $\lambda_\opt$ are indicated by horizontal lines for both pairs of parameters.}
    \label{fig:9}
\end{figure}

The observed differences between the optimal input mean photon numbers naturally lead to a difference in the corresponding single-photon probabilities. In Fig.~\ref{fig:8} we present the difference $\Delta_P^{\SPD,\lambda_n,\lambda_n^{\text{ref}}}=P_{1,\max}^{\SPD,\lambda_n}-P_{1,\max}^{\SPD,\lambda_n^{\text{ref}}}$ between the maximal single-photon probabilities for SPD obtained by assuming different input mean photon numbers $\lambda_{n,\opt}$ and input mean photon numbers $\lambda_{n,\opt}^{\text{ref}}$ of Eq.~\eqref{formula:maz}, respectively, in each of the multiplexed units, as a function of the reflection efficiency $V_r$ and the detector efficiency $V_D$ for the general transmission coefficient $V_b=0.85$. Below the continuous black line the difference $\Delta_P^{\SPD,\lambda_n,\lambda_n^{\text{ref}}}$ is higher than $10^{-3}$.
The figure shows that the unit-wise optimization of the input mean photon numbers $\lambda_n$ with the method based on the full statistical treatment as described in this paper always leads to higher single-photon probabilities than the ones that can be obtained by the optimization based on Eq.~\eqref{formula:maz}. Although for higher values of the reflection efficiency $V_r$ the two optimization technique leads to nearly equal single-photon probabilities, for lower values of $V_r$ the advantage of our procedure is relevant.

Finally, we address the problem of stability: the extent of the tolerable deviation from the optimal values of the different input mean photon numbers $\lambda_{n,\opt}$ so that the resulting single-photon probabilities $P_1$ remain higher than the maximal value that can be achieved with identical optimal input mean photon numbers $\lambda_{\opt}$.
In Fig.~\ref{fig:9} we present the optimal input mean photon numbers $\lambda_{n,\opt}$ with error bars showing the amount of allowed uniform deviation from the optimal values of $\lambda_n$, as a function of the sequential number $n$ for SPD strategy for two pairs of the reflection and detector efficiencies $V_r=0.99$, $V_D=0.9$, and $V_r=0.9$, $V_D=0.8$, and for the general transmission coefficient $V_b=0.98$.
For the reflection efficiency $V_r=0.99$ and the detector efficiency $V_D=0.9$, the maximal single-photon probability achieved with different and identical input mean photon numbers are $P_{1,\max}^{\SPD,\lambda_n}=0.9059$ and $P_{1,\max}^{\SPD,\lambda}=0.9052$, respectively, and the range defined by the maximal uniform deviation from the optimal values of $\lambda_n$ is $[-0.05,0.057]$.
For the other set of parameters, $V_D= 0.8$, $V_r= 0.9$ , the maximal single-photon probabilities are $P_{1,\max}^{\SPD,\lambda_n}=0.7281$ and $P_{1,\max}^{\SPD,\lambda}=0.7245$, and the range defined by the maximal uniform deviation from the optimal values of $\lambda_n$ is $[-0.129,0.15]$.
Based on these typical examples one can conclude that the realization of the unit-wise optimized input mean photon numbers is likely to be feasible in experiments with a precision sufficient for improving the performance of the system compared to the case of using identical optimal input mean photon numbers.

\section{Conclusion}\label{Sec:conc}
We have developed a full statistical description of multiplexed single-photon sources equipped with photon-number-resolving detectors that incorporates the use of different input mean photon numbers in each of the multiplexed units.
This theory includes all relevant loss mechanisms and enables the maximization of the single-photon probabilities at the output under realistic conditions by optimizing the different input mean photon numbers unit-wise and by determining the optimal system size.
Moreover, embedding photon-number-resolving detection in the model allows for the optimization of the detection strategy which is characterized by the set of number of detected photons for which the generated signal photons are allowed to enter the multiplexer.

Using this full statistical description we have analyzed in detail periodic single-photon sources based on asymmetric spatial multiplexing realized with general asymmetric routers and photon-number-resolving detectors for experimentally feasible loss parameters.
We have shown that the use of optimal different input mean photon numbers results in maximal single-photon probabilities higher than those that can be achieved by using optimal identical input mean photon numbers in this system.
A considerable enhancement can be achieved for threshold detection and for the $S=\{1,2\}$ detection strategy.
We have shown that this latter is the optimal detection strategy for a part of the considered ranges of parameters.  
In the case of single photon detection which is the optimal detection strategy for the remaining, bigger part of the parameters' ranges this enhancement is moderate.
However, even in the latter case the enhancement is relatively larger for suboptimal system sizes, especially for multiplexers with higher total transmission efficiencies and when detectors with lower detector efficiencies are used.
A special advantage of the unit-wise optimization of the input mean photon numbers is that it can result in the decrease of the optimal system size needed to maximize the single-photon probability.

We have compared our results to those that can be obtained by the application of a simple functional dependence for determining the different input mean photon numbers as described before in the literature. From this we can conclude that the unit-wise numerical optimization of the input mean photon numbers with our method leads to higher single-photon probabilities. The difference is significant in most parts of the parameters' ranges.

We have considered the experimental feasibility of the key ingredient of our scheme, the unit-wise optimized input mean photon numbers. It appears that these can be realized in experiments with a precision sufficient to achieve an improved performance of the system.

The most promising result of our analysis is that the highest single-photon probability is 0.935 that, in principle, can be achieved in single-photon sources based on asymmetric spatial multiplexing using state-of-the-art bulk-optical devices. To our knowledge this value is the highest one reported in any proposed scheme thus far in the literature.

\acknowledgments{
This research was supported by the National Research, Development and Innovation Office, Hungary (Projects No.\ K124351 and No.\ 2017-1.2.1-NKP-2017-00001 HunQuTech). The project has also been supported by the European Union (Grants No. EFOP-3.6.1.-16-2016-00004, No. EFOP-3.6.2-16-2017-00005 and No. EFOP-3.4.3-16-2016-00005).}


%
\end{document}